\documentclass{ifacconf}

\usepackage{graphicx}      
\usepackage{natbib}        

\usepackage{textcomp}
\usepackage{subfigure}
\usepackage{algorithm}

\usepackage{tikz}
\usetikzlibrary{shapes,arrows,positioning}
\usepackage{multirow}

\usepackage{amsmath}
\usepackage{graphicx}      
\usepackage{natbib}        
\usepackage{array}
\usepackage{algorithm}
\usepackage{etoolbox}
\let\classAND\AND
\let\AND\relax
\usepackage{algorithmic}

\let\AND\classAND
\AtBeginEnvironment{algorithmic}{\let\AND\algoAND}

\usepackage{amssymb,latexsym,amsfonts,amsmath}
\usepackage{mathrsfs}
\usepackage{graphicx} 
\usepackage{paralist}
\usepackage{dsfont}
\usepackage{eso-pic}
\usepackage{epsfig}
\usepackage{mathtools}
\usepackage{epstopdf}
\usepackage{lipsum}
\usepackage{psfrag}
\usepackage{xparse}
\usepackage{gensymb}
\usepackage{bbm}
\usepackage{mdframed,lipsum}
\usepackage{tikz}
\usetikzlibrary{calc,shapes,arrows}
\usepackage{url}

\usepackage[many]{tcolorbox}
\usetikzlibrary{calc}
\tcbuselibrary{skins}

\newtcolorbox{resp}[1][]{%
	enhanced jigsaw,%
	colback=gray!5!white,%
	colframe=gray!80!black,%
	size=small,%
	boxrule=1pt,%
	halign title=flush center,%
	coltitle=black,%
	breakable,%
	drop shadow=black!50!white,%
	attach boxed title to top left={xshift=1cm,yshift=-\tcboxedtitleheight/2,yshifttext=-\tcboxedtitleheight/2},%
	minipage boxed title=3cm,%
	boxed title style={%
		colback=white,%
		size=fbox,%
		boxrule=1pt,%
		boxsep=2pt,%
		underlay={%
			\coordinate (dotA) at ($(interior.west) + (-0.5pt,0)$);
			\coordinate (dotB) at ($(interior.east) + (0.5pt,0)$);
			\begin{scope}[gray!80!black]
				\fill (dotA) circle (2pt);
				\fill (dotB) circle (2pt);
			\end{scope}
		}%
	},%
	#1%
}

\newtheorem{theorem}{Theorem}[section]
\newtheorem{lemma}[theorem]{Lemma}
\newtheorem{problem}[theorem]{Problem}

\newtheorem{definition}[theorem]{Definition}

\newtheorem{remark}[theorem]{Remark}
\newtheorem{assumption}[theorem]{Assumption}
\numberwithin{equation}{section}
\usepackage{mathtools}

\usepackage{xspace}

\newcommand{\R}{{\mathbb{R}}}

\newcommand{\N}{{\mathbb{N}}}

\newcommand{\EE}{\mathds{E}}

\newcommand{\pr}{\mathds{P}}

\begin{document}
\begin{frontmatter}

\title{Data-Driven Safety Verification of Stochastic Systems via Barrier Certificates}

\thanks{This work was supported in part by the H2020 ERC Starting Grant AutoCPS (grant agreement No. 804639) and by the EPSRC-funded CodeCPS project (EP/V043676/1).}

\author[First]{Ali Salamati} 
\author[Second]{Abolfazl Lavaei} 
\author[Third]{Sadegh Soudjani}
\author[Fourth,First]{Majid Zamani}

\address[First]{Department of Computer Science, Ludwig-Maximilians-Universit\"{a}t M\"{u}nchen, Germany, (e-mail: ali.salamati@lmu.de)}
\address[Second]{Institute for Dynamic Systems and Control, ETH Zurich, Switzerland, (e-mail: alavaei@ethz.ch)}
\address[Third]{School of Computing, Newcastle University, United Kingdom, (e-mail: sadegh.soudjani@newcastle.ac.uk)}
\address[Fourth]{Department of Computer Science, University of Colorado Boulder, the USA, (e-mail: majid.zamani@colorado.edu)}

\begin{abstract}
In this paper, we propose a data-driven approach to formally verify the safety of (potentially) unknown discrete-time continuous-space stochastic systems. The proposed framework is based on a notion of barrier certificates together with data collected from trajectories of unknown systems. We first reformulate the barrier-based safety verification as a robust convex problem (RCP). Solving the acquired RCP is hard in general because not only the state of the system lives in a continuous set, but also and more problematic, the unknown model appears in one of the constraints of RCP. Instead, we leverage a finite number of data, and accordingly, the RCP is casted as a scenario convex problem (SCP). We then relate the optimizer of the SCP to that of the RCP, and consequently, we provide a safety guarantee over the unknown stochastic system with a priori guaranteed confidence. We apply our approach to an unknown room temperature system by collecting sampled data from trajectories of the system and verify formally that temperature of the room lies in a comfort zone for a finite time horizon with a desired confidence.
\end{abstract}


\end{frontmatter}

\section{introduction}
Stochastic dynamical systems have gained remarkable attentions as an important modeling framework characterizing many engineering systems; they play
crucial roles in real-life safety-critical applications, in which system’s failures (\emph{e.g.,} collision) are not
acceptable. Examples of such applications include
traffic networks, self-driving cars, and so on. Formal analysis of this type of complex systems against some high-level
specifications, \emph{e.g.,} those expressed as linear temporal logic
(LTL) formulae \citep{kesten1998algorithmic}, is inherently very challenging due to uncountable sets of states and actions together with uncertainties inside dynamics. To mitigate this complexity, abstraction-based techniques have been studied for verification and synthesis of stochastic dynamical systems \citep{LAB15,SSoudjani2014,SVORENOVA2017230,azuma2014discrete,haesaert2020formal,majumdar2021symbolic,Lavaei_Survey}.
To make these techniques scalable, other approaches based on adaptive gridding \citep{esmaeil2013adaptive}, and compositional abstraction-based methods \citep{soudjani2015dynamic,lavaei2018CDCJ,lavaei2018ADHSJ,lavaei2019Thesis} have been introduced in the relevant literature to efficiently handle verification and synthesis problems on such classes of models. 

Another promising approach for safety verification of nonlinear stochastic systems is using \emph{barrier certificate} introduced by \cite{prajna2004safety}.
This approach has received significant attentions in the past decade, as a \emph{discretization-free approach}, for formal verification and synthesis of non-stochastic \citep{borrmann2015control,wang2017safety}, and stochastic dynamical systems \citep{prajna2007framework,zhang2010safety,yang2020efficient}. However, in all the aforementioned works, one needs to know precise models of dynamical systems to construct those barrier certificates, and accordingly, those approaches are not applicable when the model is (partially) unknown. Therefore, data-driven methods are essential to directly collect data from the systems for their formal analysis.
In this regard, a framework is proposed by \cite{sadraddini2018formal} to use input-output data for an unknown system to synthesize controllers from signal temporal logic specifications by finding a set-valued piecewise affine model that contains all the possible behaviors of the original system. A data-driven approach for the formal verification of partially unknown stochastic system against signal temporal logic properties is recently proposed by \cite{salamati2020data}. Reinforcement learning (RL) schemes to synthesize policies for unknown continuous-space stochastic systems are proposed by~\cite{lavaei2020ICCPS} and by \cite{kazemi2020formal} while providing convergence to near-optimal policies. An optimization-based approach is suggested by \cite{robey2020learning} to learn a control barrier certificate through safe trajectories under suitable Lipschitz smoothness assumptions on the dynamical systems.

There have also been some works in the setting of robust optimization problems using \emph{scenario-based} approaches. A  probabilistic framework based on scenario approach for providing a bound on the number of required samples to obtain a priori specified level of guarantee of robustness is proposed by \cite{calafiore2006scenario}. Worst-case violation of sampled convex programs is investigated by \cite{kanamori2012worst}. A novel framework for establishing a relation between the optimal value of a scenario convex problem and that of the original robust linear programming and its extension to a certain class of non-convex problems is proposed by \cite{esfahani2014performance}. A technique for solving chance-constrained optimizations is proposed by \cite{SM18_Concentration} that does not require any convexity assumption but utilizes concentration properties of the underlying probability distributions.

Our main contribution in this work is to develop a data-driven approach to formally verify the safety of (potentially) unknown discrete-time continuous-space stochastic systems. We first cast the barrier-based safety problem as a robust convex problem (RCP). Since solving the acquired RCP is not possible due to the unknown model that appears in one of the constraints, we propose the scenario convex problem (SCP) corresponding to the original RCP by employing a finite number of data collected from the system. Then inspired by \cite{esfahani2014performance}, we make a bridge between the optimizer of the SCP to that of the RCP, and accordingly, we provide a safety guarantee over the unknown stochastic system with a priori guaranteed confidence. We finally apply our approaches to an unknown room temperature system. Proofs of all statements are omitted here due to lack of space.

\section{System Definition and Problem Statement}
\label{sec:problemset}
\subsection{Notations and Preliminaries}
The set of positive integers, non-negative integers, real numbers, non-negative real numbers, and positive real numbers are denoted by $\mathbb{N} := \{1,2,3,\ldots\}$, $\mathbb{N}_0 := \{0,1,2,\ldots\}$, $\mathbb{R}$, $\mathbb{R}_0^+$ , and $\mathbb{R}^+$, respectively. We denote the indicator function by $\mathbbm{1}_\mathcal A(X):~X\rightarrow \{0,1\}$, where $\mathbbm{1}_\mathcal A(x)$ is $1$ if and only if $x\in \mathcal A$, and $0$ otherwise. We denote by $\Vert x\Vert$ the Euclidean norm of $x\in\R^n$. We also denote by $\|A\|_F$ the \emph{Frobenius} norm of any matrix $A\in\mathbb R^{m\times n}$. The absolute value of a real number $x$ is denoted by $|x|$.
Given a symmetric matrix $A$, $\lambda_{\max}(A)$ denotes the maximum eigenvalue of $A$. Given $N$ vectors $x_i \in \mathbb R^{n_i}$, $n_i\in \mathbb N_{\ge 1}$, and $i\in\{1,\ldots,N\}$, we use $x = [x_1;\ldots;x_N]$ to denote the corresponding column vector of the dimension $\sum_i n_i$. Considering a random variable $\mathrm{z}$, $\text{Var}(\mathrm{z})$ denotes its variance. If a system, represented by $\mathcal{S}$, satisfies a property $\Psi$, it is denoted by $\mathcal{S} \models \Psi$. We also use $\models$ in this paper to show the feasibility of a solution for an optimization problem.

The measurable space $\mathfrak{B}(X)$ is a Borel $\sigma$-algebra on the state space X denoted by $(X,\mathfrak{B}(X))$ and the sample space is denoted by $\Omega$. We have two probability spaces in this work. The first one is represented by $(X,\mathfrak{B}(X),\pr)$ which is the probability space defined over the state space $X$ with $\pr$ as a probability measure. The second one, $(V_w,\mathfrak{B}(V_w),\pr_w)$, defines the probability space over $V_w$ which is the set of independent and identically distributed (i.i.d.) random variables $w$ with $\pr_w$ as its probability measure.

\subsection{System Definition}
In this work, we deal with unknown discrete-time stochastic systems as formalized in the next definition.
\begin{definition}
	Consider a discrete-time stochastic system (dt-SS), denoted by $\mathcal{S} = (X,V_w,w,f)$, represented as the following:
	\begin{equation} 
		\mathcal{S}\!: x(t+1)=f(x(t),w(t)), \quad t\in\mathbb{N}_{0},
		\label{eq:mainsystem}
	\end{equation}
	where $X$ and $V_w$ are Borel $\sigma$-algebra on the state space $\mathbb{R}^n$ and uncertainty spaces, respectively. Variable $x$ denotes state of the system as $x:=\{x(t):\Omega \rightarrow X, t\in\mathbb{N}_0\}$, and variable $w$ introduces a sequence of i.i.d random variables on the Borel space $V_w$ and it is expressed as $w:=\{w(t):\Omega \rightarrow V_w, t\in\mathbb{N}_0\}$. Map $f:X\times V_w\rightarrow X$ is a measurable function characterizing the state evolution of the system. A finite trajectory of the system in \eqref{eq:mainsystem} is denoted by $\xi(t):=x(0)x(1)\ldots x(t) , t\in \mathbb{N}_0$.
	\label{def:mainsys}
\end{definition}

\subsection{Problem Statement}
\label{sub:ps}
In this work, we assume that the map $f$ and the distribution of the stochasticity $\pr_w$ are unknown. Instead, we only observe $N$ i.i.d. sampled data collected randomly with a uniform distribution from $X$ as
\begin{equation}
	\mathcal{W}_N:=\{\hat{x}_i,1\leq i \leq N\}\subseteq X. 
	\label{eq:data}
\end{equation}

Next definition provides the safety specification for the unknown stochastic system in Definition~\ref{def:mainsys}.
\begin{definition}\label{safe}
	Given a safety specification $\Psi$, the system $\mathcal{S}$ is called safe for a finite time horizon $\mathcal{T}_h \in \mathbb{N}_0$, denoted by $\mathcal{S}\models_{\mathcal{T}_h} \Psi,$ if all trajectories of $\mathcal{S}$ started from an initial set $X_{in}\subseteq X$ never reach an unsafe set $X_u\subseteq X$.
\end{definition}

Since we do not have any knowledge about the model and the distribution of the noise, the question of interest here is that: ``\emph{can one judge about the safety of a stochastic system only by leveraging  data collected from trajectories of the unknown system?}" This inspiring question can be formalized as the following problem.
\vspace{0.2cm}
\begin{resp}
	\begin{problem}
		\label{prob:problem}
		Consider a potentially unknown stochastic system $\mathcal{S}$ as in Definition~\ref{def:mainsys}. Given $N$ sampled data as in \eqref{eq:data}, provide a formal guarantee on the satisfaction of the safety specification $\Psi$ with a priori probability lower bound $1-\rho$, $\rho\in (0,1]$, \emph{i.e.,}
		\begin{equation*}
			\pr_w\big(\mathcal{S}\models_{\mathcal{T}_h} \Psi\big)\ge 1-\rho.
		\end{equation*} 
	\end{problem} 
\end{resp}
\vspace{0.2cm}

To address this problem, we first present the safety analysis of stochastic systems via barrier certificates as in the next section.

\section{Barrier Certificates}\label{sec:barrier}
In this section, we state an existing result in the literature that uses the notion of barrier certificate (BC) to compute a lower bound on the probability of satisfying safety specifications for discrete-time stochastic systems. Let us start by formally defining barrier certificates.

\begin{definition}
	Given a stochastic system $\mathcal{S}$ in Definition~\ref{def:mainsys}, a nonnegative function $\mathrm{B}:X\rightarrow\mathbb{R}_{0}^{+}$ is called a barrier certificate (BC) for $\mathcal{S}$ if there exist constants $\lambda > 1,\;\text{and}\;c\in \mathbb{R}_{0}^{+}$ such that
	\begin{align}
		&\mathrm{B}(x)\leq 1,\qquad\qquad\qquad\qquad\quad~~~~~\;\forall x\in X_{in},	\label{eq:bar1}\\
		&\mathrm{B}(x)\geq \lambda, \qquad\qquad\qquad\qquad\qquad~~~\!\forall x\in X_u,\label{eq:bar2}\\
		&\EE\Big[\mathrm{B}(f(x,w))\mid x\Big]\leq~\;\mathrm{B}(x)+c,\;\;\quad\!\forall x \in X,\label{eq:bar3}
	\end{align}
	where $X_{in},X_u\subseteq X$ are initial and unsafe sets corresponding to a given safety specification $\Psi$, respectively. 
	\label{def:barrier}
\end{definition}

Next theorem, borrowed from \citep{jagtap2019formal}, provides a lower bound on the probability of satisfaction of the safety specification for a dt-SS.
\vspace{.2cm}
	\begin{theorem}
		\label{theo:kushner}
		Consider a stochastic system $\mathcal{S}$ as defined in Definition~\ref{def:mainsys}, and a safety specification $\Psi$. Suppose that there exists a non-negative barrier certificate $\mathrm{B}$ satisfying conditions \eqref{eq:bar1}-\eqref{eq:bar3}. Then
		\begin{equation*}
			\pr\big(\mathcal{S}\models_{\mathcal{T}_h} \Psi\big)\ge 1-\frac{1+c\;\mathcal{T}_h}{\lambda}.
		\end{equation*}
		\label{theo:barrier1}
	\end{theorem}
In this work, we resort to find barrier certificates $\mathrm{B}(b,x)$ which are polynomial functions of $x$ with coefficients stored in $b$. Such a polynomial with degree $k\in \mathbb{N}$ can be written as
\begin{align*}
	\mathrm{B}(b,x)=\sum_{(\iota)}b_{(\iota)} x_1^{\iota_1}\ldots x_n^{\iota_n},
\end{align*}
where the sum is over all possible $(\iota) := (\iota_1,\ldots,\iota_n)$ with $\iota_1,\ldots,\iota_n\ge 0$ and $\iota_1+\ldots+\iota_n = k$.
Finding barrier certificates then boils down to determining their coefficients, namely $b_{(\iota)}$. In the next section, we propose our data-driven scheme for the construction of barrier certificates from data collected from trajectories of unknown systems.

\section{Data-driven Safety Problem}
\label{sec:datadriven}
In this section, we first cast the barrier-based safety problem into a robust convex program (RCP).
In particular, direct use of Theorem~\ref{theo:kushner} requires solving an RCP formulated as
\begin{align} 
	\textbf{RCP}:\left\{
	\begin{array}{ll}
		\underset{d}{\textbf{min}}\quad\; \mathcal{K}\quad \quad\\
		\textbf{s.t.}\quad\;
		\max\Big\{g_z(x,d)\Big\} \!\leq\! 0, z \!\in\!\{1,\dots,5\},\forall x \!\in\! X,\\
		\qquad ~~ \;d=[\mathcal{K};\lambda;c;b_{(\iota)}],\\
		\qquad ~~~c \in \mathbb{R}_{0}^{+}, \; \lambda > 1,~~\mathcal{K}\in \mathbb{R},
	\end{array}
	\right.
	\label{eq:RCP}
\end{align}
where,
\begin{align}
	&g_1(x,d)=-\mathrm{B}(b,x)-\mathcal{K},\nonumber\\
	&g_2(x,d)=(\mathrm{B}(b,x)-1)\mathbbm{1}_{X_{in}}(x)-\mathcal{K}, \nonumber\\
	&g_3(x,d)=(-\mathrm{B}(b,x)+\lambda)\mathbbm{1}_{X_{u}}(x)-\mathcal{K},\nonumber\\
	&g_4(x,d)=\frac{1+c\;\mathcal{T}_h}{\rho}-
		\lambda-\mu-\mathcal{K},\nonumber\\
	&g_5(x,d)=\EE\Big[\mathrm{B}(b,f(x,w))\mid x\Big]-\mathrm{B}(b,x)-c-\mathcal{K},
	\label{eq:sampledcondition}
\end{align}
where $\mu < 0$ and $1-\rho$ with $\rho \in (0,1]$ is a priori lower bound for the probability of satisfaction as in Problem \ref{prob:problem}. It is not hard to verify that the RCP in~\eqref{eq:RCP} always has a feasible solution. For instance as a trivial solution, by choosing $\lambda=2$, $c=0$, $\mu = -1$, and coefficients of $\mathrm{B}(b,x)$ to be zero, there exists always a large enough $\mathcal{K}$ such that $\frac{1}{\rho}-1<\mathcal{K}$. The obtained barrier certificate by solving this RCP always satisfies conditions \eqref{eq:bar1}-\eqref{eq:bar3} for non-positive values of $\mathcal{K}$.

Finding an optimal solution for the RCP in \eqref{eq:RCP} is hard in general because not only there is no access to the model of system (\emph{i.e.,} $f$), but also the state of the system lives in the continuous set $X$. To tackle this problem, we collect data from trajectories of unknown systems and propose a corresponding scenario convex program of RCP, denoted by SCP\textsubscript{N}, as the following:
	\begin{align} 
		\textbf{SCP\textsubscript{N}}:\left\{
		\begin{array}{ll}
			\underset{d}{\textbf{min}}\quad \mathcal{K}\quad \quad\\
			\textbf{s.t.}\quad
			\max\Big\{g_z(\hat{x}_i,d)\Big\} \!\leq\! 0, z \!\in\!\{1,\dots,5\},\\
			\quad \quad ~\;\;\forall \hat{x}_i \in X, \forall i\in \{1,\ldots,N\},\\
			\qquad ~~~d=[\mathcal{K};\lambda;c;b_{(\iota)}],\\
			\qquad ~~~c \in \mathbb{R}_{0}^{+},  \lambda >1,~\mathcal{K}\in \mathbb{R}.
		\end{array}
		\right.
		\label{eq:SCN1}
	\end{align}
Since there is no closed-form solution for the expected value in $g_5$, we instead use empirical approximation and propose a new scenario convex problem, denoted by SCP\textsubscript{w} as follows:
	\begin{align} 
		\textbf{SCP\textsubscript{w}}\!:\!\left\{
		\begin{array}{ll}
			\underset{d}{\textbf{min}}\quad \mathcal{K}\quad \quad\\
			\textbf{s.t.}\quad\;
			\max\!\Big\{\!g_z(\hat{x}_i,d),\bar{g}_5(\hat{x}_i,d)\!\Big\} \!\!\leq\! 0,z \!\in\!\!\{1,\dots,4\},\quad \\
			\quad\quad~~~ \forall \hat{x}_i \in X, \forall i\in\{1,\ldots,N\},\\
			\qquad ~~~d=[\mathcal{K};\lambda;c;b_{(\iota)}],\\
			\qquad ~~~c \in \mathbb{R}_{0}^{+},\;  \lambda >1, ~\mathcal{K}\in \mathbb{R},
		\end{array}
		\right.
		\label{eq:SCN2}
	\end{align}
	with
	\begin{align}
		\bar{g}_5(\hat{x}_i,d)=\frac{1}{\hat{N}}\sum_{j=1}^{\hat{N}}\mathrm{B}(b,f(\hat{x}_i,\hat{w}_j))-  \;\mathrm{B}(b,\hat{x}_i)-c+\delta-\mathcal{K},
		\label{eq:empirical}
	\end{align}
where $\hat{N}\in\mathbb{N}$ and $\delta\in \mathbb{R}^+$ are respectively the number of samples required for the empirical approximation, and the error introduced by this approximation. The optimal value for the objective function of SCP\textsubscript{w} is denoted by $\mathcal{K}^*(\mathcal{W}_N)$. We also denote by $\hat{\mathrm{B}}(b,x)$ the barrier function constructed based on the solution of the scenario problem in \eqref{eq:SCN2}. 

In the new scenario problem, $f(\hat{x}_i,\hat{w}_j)$ is the realization of the unknown system started from an initial state $\hat{x}_i$ for a noise realization $\hat{w}_j$. The empirical approximation for each sample $\hat{x}_i$ is computed over $\hat{N}$ different realizations of noise $\hat{w}_j,j\in\{1,\dots,\hat{N}\}$. This approximation introduces an error in $\bar g_5$, represented by $\delta$, which makes it more conservative. We use Chebyshev's inequality~\citep{hernandez2001chebyshev} to quantify the error by providing a probabilistic upper bound for it. To do so, we need to define 
variance of the empirical approximation, denoted by $\sigma^2$, as follows:
\begin{align}
	\sigma^2 := \text{Var}\Big(\frac{1}{\hat{N}}\sum_{j=1}^{\hat{N}}\mathrm{B}(b,f(x,\hat{w}_j))\Big).
	\label{eq:variance}
\end{align}  

Next theorem shows that the barrier certificate characterized by the optimal solution of SCP\textsubscript{w} is a feasible BC for SCP\textsubscript{N} in \eqref{eq:SCN1} with some certain confidence.

\begin{theorem}
	Suppose that $\hat{\mathrm{B}}(b,x)$ is a solution of SCP\textsubscript{w}. Then for a priori value of the error $\delta\in \mathbb{R}^+$, a desired stochastic confidence $\beta_s \in (0,1]$, and a given upper bound $\hat{M}$ on the variance of the barrier certificate applied on $f$, \emph{i.e.,} $\text{Var}\big(\mathrm{B}(b,f(x,w))\big)\leq \hat{M}\in \mathbb{R}^+$, one has
	\begin{align}
		\pr_w\Big(\hat{\mathrm{B}}(b,x) \models\text{SCP\textsubscript{N}}\Big)\geq 1- \beta_s,
		\label{eq:feasibility}
	\end{align}
	provided that
	$\hat{N} \geq \frac{\hat{M}}{\delta^2\beta_s}$.
	\label{theo:empirical}
\end{theorem}

\begin{remark}
	When the underlying system is affected by an additive noise, \emph{i.e.,} 
	\begin{align}
		x(t+1) = f_a(x(t)) + w(t),\nonumber
	\end{align}
	the condition $\text{Var}\big(\mathrm{B}(b,f(x,w))\big)\leq \hat{M}\in \mathbb{R}^+$ boils down to having a bounded $f_a(x(t))$, $\forall t\in\N_0$. In this case, the value of $\hat{M}$ is computable using a bound on $f_a(x(t))$ and bounds on moments of $w$.
	For instance, in the case of one-dimensional systems ($n=1$), we have $\mathrm{B}(b,x) = \sum_{\iota=0}^{k} b_{\iota} x^{\iota}$ and the variance of $\mathrm{B}(\cdot)$ can be expanded as follows:
	\begin{align*}
		& \text{Var}(b,\mathrm{B}(f(x,w))) = \text{Var}\Big(\sum_{\iota=0}^{k} b_{\iota} f(x,w)^{\iota}\Big)\\
		&=\!\text{Var}\Big(\!\sum_{\iota=0}^{k} b_{\iota} (f_a(x) + w)^{\iota}\Big) \!=\!\text{Var}\Big(\!\sum_{\iota}^{k}\!\sum_{j=0}^{\iota}\! b_{\iota} {\iota \choose j} \!f_a(x)^{\iota-j} w^{j}\!\Big)\\
		& =\!\text{Var}\Big(\sum_{j=1}^{k}\mathrm{g}_{j}(x) w^{j}\Big) \text{ with } \mathrm{g}_{j}(x) := \sum_{\iota=j}^{k} b_{\iota} {\iota \choose j} f_a(x)^{\iota-j}\\
		& =\! \sum_{j=1}^{k}\sum_{z=1}^{k} \mathrm{g}_j(x)\mathrm{g}_z(x)(\EE[w^{j+z}] - \EE[w^j]\EE[w^z]).
	\end{align*}
	This means the variance can be bounded using upper bounds of $f_a(\cdot)$ and moments of $w$.
\end{remark}

As seen from Theorem~\ref{theo:empirical}, higher number of samples  $\hat{N}$ is needed in order to have a smaller empirical approximation error $\delta$, and accordingly, provide a better confidence bound. In fact, $\hat{N}$ and $\delta$ are required to solve SCP\textsubscript{w} in \eqref{eq:SCN2}. Later in the next section, we show how the value of  $\beta_s$ affects the total confidence concerning the safety of the stochastic system. 

\section{Safety Guarantee over Unknown Stochastic Systems}\label{sub:main_safe}
In the previous section, we showed that using a finite number of data, the original RCP can be corresponded to SCP\textsubscript{N} for which the solution can be approximated with an arbitrary precision (cf. Theorem~\ref{theo:empirical}). In this section, we establish the missing connection between solutions to the original RCP and the corresponding SCP\textsubscript{N} by employing the fundamental results in \citep{esfahani2014performance}. Consequently, we provide a safety guarantee over the unknown stochastic system with a priori guaranteed confidence. Before providing the main result, we need to raise the following assumption.
\begin{assumption}
	Suppose functions $g_1$, $g_2$, $g_3$ and $g_5$ are all Lipschitz continuous with respect to $x$ with Lipschitz constants $\mathrm{L}_{x_1}$, $\mathrm{L}_{x_2}$, $\mathrm{L}_{x_3}$, and $\mathrm{L}_{x_5}$, respectively.    
	\label{assum:lip}
\end{assumption}
We utilize Assumption~\ref{assum:lip} and propose the next theorem that establishes
a relation between the optimal values of SCP\textsubscript{w} and that
of the original RCP, and accordingly, verify the safety of unknown stochastic systems with a priori guaranteed confidence.
\vspace{.25cm}

\begin{theorem}
	Consider an unknown dt-SS as in~\eqref{eq:mainsystem}, and initial and unsafe regions $X_{in}$ and $X_u$, respectively. 
		Let Assumption~\ref{assum:lip} hold. Consider the corresponding SCP\textsubscript{w} as in~\eqref{eq:empirical} with its associated optimal value $\mathcal{K}^*(\mathcal{W}_N)$ and solution $d^* = [\mathcal K^*;\lambda^*;c^*;b^*_{\iota_1,\ldots,\iota_k}]$, with
		$\hat{N}$ as in Theorem~\ref{theo:empirical} and $N\geq N\big(\bar{\epsilon},\beta\big)$, where 
		\begin{equation}
			\label{eq:number_samples}
			N(\bar{\epsilon},\beta) := \min \Big\{N\in\mathbb{N}\mid\sum_{i=0}^{\mathcal{\mathcal{Q}}+2} \dbinom{N}{i}\bar{\epsilon}^{\;i}
			(1-\bar{\epsilon})^{N-i} \leq \beta \Big\},
		\end{equation}
		$\epsilon,\beta \in [0,1]$ with $\epsilon \leq \mathrm{L}_{x} := \max \big\{\mathrm{L}_{x_1},\mathrm{L}_{x_2},\mathrm{L}_{x_3},\mathrm{L}_{x_5}\big\}$, $\bar{\epsilon}:=(\frac{\epsilon}{\mathrm{L}_{x}})^n$, and $\mathcal{Q}$ is the number of coefficients of barrier certificate. Then the following statement holds with a confidence of at least $1-\beta-\beta_s$, with $\beta_s \in (0,1]$ as in Theorem~\ref{theo:empirical}: For a given $\rho \in (0,1]$, if $\mathcal{K}^*(\mathcal{W}_N)+\epsilon \leq0$ then 
		\begin{align}
			&\pr_w\big(\mathcal{S}\models_{\mathcal{T}_h} \Psi \big)\ge1-\rho.
			\label{RCC}
	\end{align}
	\label{theo:peyman}
\end{theorem}

\vspace{-5mm}
\begin{remark}
	Theorem \ref{theo:peyman} establishes a connection between
	the optimal value of SCP\textsubscript{w} and that of the original RCP in \eqref{eq:RCP}, and as a result, provides a lower bound on the satisfaction probability of safety specifications for the unknown stochastic system with a confidence of at least $1-\beta-\beta_s$. According to \citep[Lemma 3.2]{esfahani2014performance}, if one makes the constraints of SCP\textsubscript{w} more negative in the amount of $\mathrm{L}_{x}\epsilon^{\frac{1}{n}}$, the constructed barrier certificate via data is a BC for the unknown system with a confidence of at least $1-\beta_s$, \emph{i.e.,} $\beta = 0$.
	\label{rem:barrier}
\end{remark}
For the sake of clarity, we present the required steps for employing Theorem~\ref{theo:peyman} in Algorithm~\ref{alg:verification}.

\begin{algorithm}[h]
	\caption{Safety verification of a potentially unknown stochastic system using collected data}
	\label{alg:verification}
	\begin{center}
		\begin{algorithmic}[1]
			\REQUIRE 
			$\beta \in[0,1]$, $\beta_s\in (0,1]$, $\rho \in(0,1]$, $\delta \in \mathbb{R}^+$, $\mathrm{L}_x \in \mathbb{R}^+$, $\hat{M} \in \mathbb{R}^+$, and degree of barrier certificate
			\STATE
			Choose $\epsilon \in [0,1]$ such that $\epsilon \leq \mathrm{L}_{x}$				
			\STATE
			Compute the minimum number of samples as  $N(\bar{\epsilon},\beta)$ according to \eqref{eq:number_samples}
			\STATE
			Compute the number of samples $\hat{N}$ required for empirical approximation according to Theorem~\ref{theo:empirical}
			\STATE
			Solve the SCP\textsubscript{w} in \eqref{eq:SCN2} with the acquired number of data $N$,  $\hat{N}$, and obtain $\mathcal{K}^*(\mathcal{W}_N)$
			\ENSURE
			If  $\mathcal{K}^*(\mathcal{W}_N)+\epsilon \leq 0 $, $\pr_w(\mathcal{S}\models_{\mathcal{T}_h} \Psi)\ge1-\rho$ with a confidence of at least $1-\beta-\beta_{s}$
		\end{algorithmic}
	\end{center}
\end{algorithm}

As it can be observed, one needs Lipschitz constant $\mathrm{L}_x$ in order to employ the proposed algorithm. In the following, we provide a systematic approach to compute the required Lipschitz constant under some assumptions. To do this, we assume that the barrier certificate is in a quadratic form.
\begin{lemma}
	Consider a nonlinear system in Definition~\ref{def:mainsys} affected by an additive Gaussian noise with zero mean and variance of $\sigma_w^2$ as the following:
	\begin{align}
		x(t+1)=f_a(x(t))+w(t).
		\label{eq:additive}
	\end{align}
	Let $\Vert f_a(x)\Vert\leq L\Vert x\Vert, L\in \mathbb{R}^+$, and $\Vert\nabla f_a(x)\Vert_F\leq \hat{L}\in \mathbb{R}^+$ where $\nabla f_a(x)$ is the gradient of $f_a(x)$. Given a quadratic barrier certificate $x^T\mathrm{P}x$ with a positive-definite matrix $\mathrm{P}$, the Lipschitz constant $\mathrm{L}_{x}$ can be upper-bounded by  $2m\;\lambda_{\max}(\mathrm{P})(L\hat{L}+1)$ in which $\Vert x\Vert\leq m \in \mathbb{R}^+$. 
	\label{lemma:nonlinearlemma}
\end{lemma}

\begin{remark}
	If the underlying dynamics in Definition~\ref{def:mainsys} is linear in the form of $x(t+1)=Ax(t)+w(t)$ with $A\in \mathbb R^{n\times n}$ and  $\Vert A\Vert_F\leq\mathcal{L}\in\mathbb{R}^+$, one can employ a similar argument as in Lemma~\ref{lemma:nonlinearlemma} and compute an upper bound for the Lipschitz constant $\mathrm{L}_{x}$ as $2m\;\lambda_{\max}(\mathrm{P})(\mathcal{L}^2+1)$.
\end{remark} 

\section{Case Study}
\label{sec:case_study}
Consider a room temperature regulation characterized by the following discrete-time stochastic control system:
\begin{align}
	\mathcal{S}\!:\;x(t+1) = &\;x(t) + \tau_s\big(\alpha_e(T_e-x(t)) + \nonumber\\&\alpha_H(T_{h}-x(t))u(t)\big) + \sigma_w w(t),
	\label{eq:sde}
\end{align}
where $x(t)$ is the temperature of the room, $u(t)$ denotes the heater valve opening as the input of the system, and $w(t)$ is a Gaussian noise with zero mean and standard deviation of $\sigma_w = 0.0125$. Moreover, $ T_h= 55 \degree C$ is the heater temperature, $T_e = 15 \degree C$ is the ambient temperature, and $\alpha_e = 8\times 10^{-3}$ and $\alpha_H =3.6\times 10^{-3}$ are heat exchange coefficients of room-ambient and room-heater, respectively. The model is adapted from \citep{girard2016safety} discretized by $\tau_s=5$ minutes. Let us consider the regions of interest as $X_{in} = [17\degree C,18\degree C]$, $X_{u} = [28\degree C,30\degree C]$, and $X = [17\degree C,30\degree C]$. We assume the model of the system is unknown. We employed the controller in \citep{jagtap2019formal} which is characterized as:
\begin{align}
	u(x) = &-1.018\times 10^{-6}x^4 + 7.563\times 10^{-5}x^3 - 0.001872x^2 \nonumber \\ &+0.02022x + .3944.
	\label{eq:cont}
\end{align}

\begin{figure}[h]
	\centering 
	\includegraphics[width=1\linewidth]{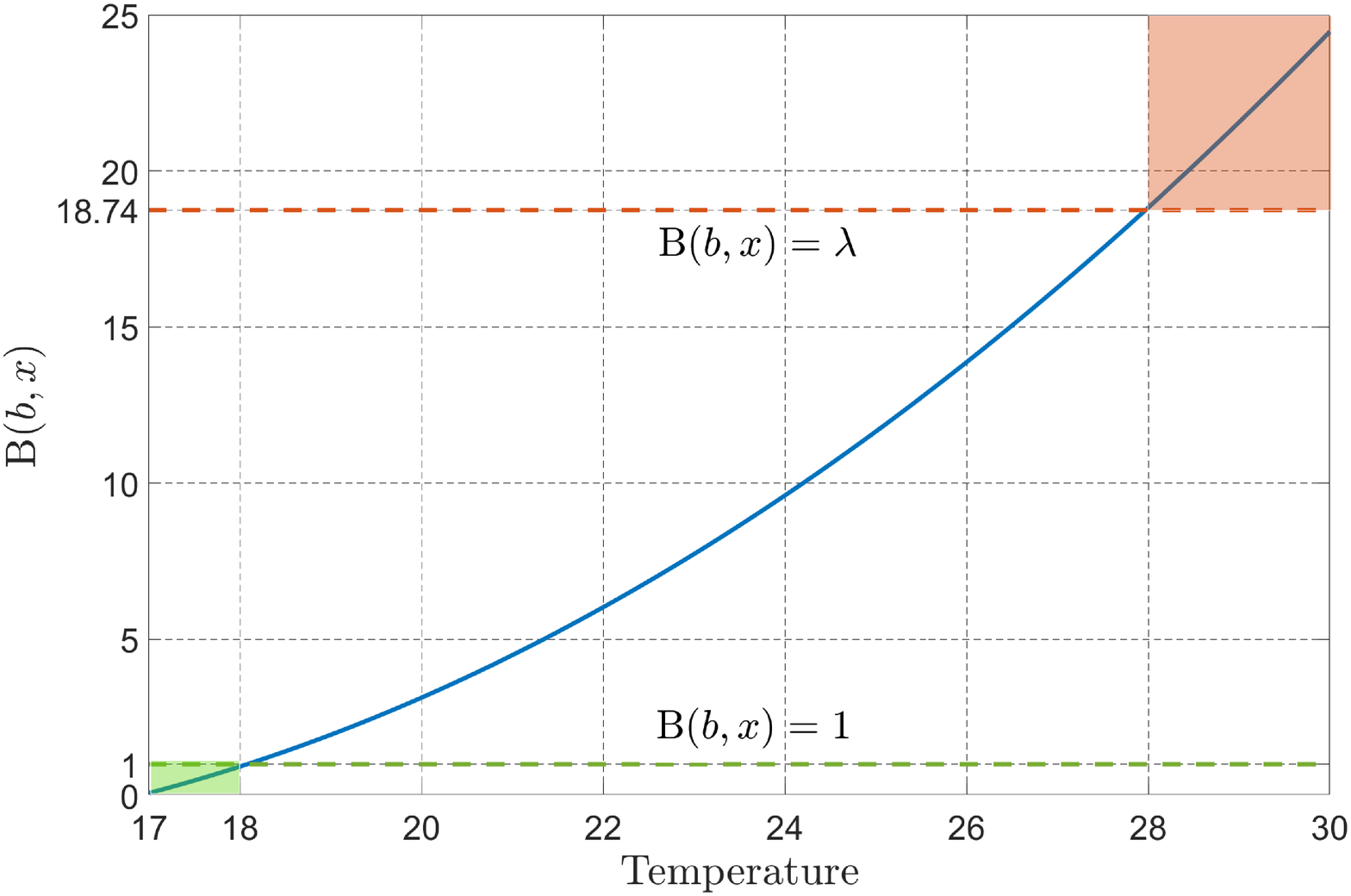}
	\caption{Barrier certificate of unknown room temperature model. Green and orange dashed lines are level sets of barrier certificate corresponding to initial and unsafe sets, respectively. According to green and orange boxes, conditions \eqref{eq:bar1} and \eqref{eq:bar2} are satisfied.}
	\label{fig:Barrier_ver1}
\end{figure}
\begin{figure}[h]
	\centering 
	\includegraphics[width=0.9\linewidth]{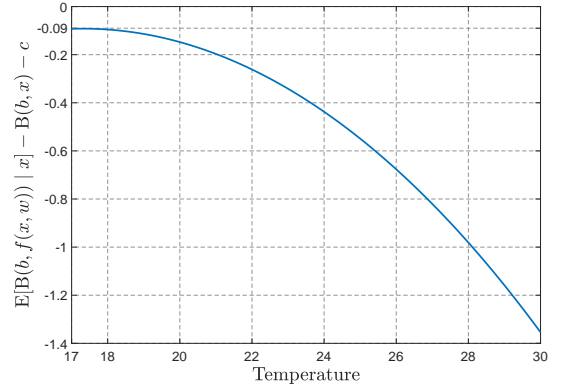}
	\caption{Satisfaction of condition \eqref{eq:bar3} in Definition~\ref{def:barrier}.}
	\label{fig:Barrier_ver2}
\end{figure}
The main goal is to verify that the temperature of the closed-loop system remains in the safe zone $[17,28]$ for the time horizon $\mathcal{T}_h=3$ (\emph{i.e.,} $15$ minutes) with some guaranteed confidence. Let us fix a barrier certificate with degree $k=2$ in the polynomial form as $x^T\mathrm{P}x=b_0x^2+b_1x+b_2$ with $b_0, b_1,b_2 \in \mathbb{R}$ where $\mathrm{P}=[b_0,\frac{b_1}{2};\frac{b_1}{2},b_2]$. According to Algorithm \ref{alg:verification}, we first choose the desired confidences $\beta,\beta_s$ as $0.005$. We also select the approximation error $\delta=0.015$. Since substituting the controller \eqref{eq:cont} in dynamics \eqref{eq:sde} results in a nonlinear dynamic, we employ Lemma \ref{lemma:nonlinearlemma} in order to compute the Lipschitz constant $\mathrm{L}_{x}$. By having $\Vert x\Vert\leq m=30$, $L\leq 2$, $\hat{L}\leq 1$, and enforcing $\lambda_{\max}(\mathrm{P})\leq12$, the Lipschitz constant can be computed as $2160$. By fixing $\epsilon = 0.03$, $\bar{\epsilon}$ can be computed as $\frac{\epsilon}{\mathrm{L}_{x}}=1.389\times10^{-5}$. Now the minimum number of samples needed to solve the SCP\textsubscript{w} in \eqref{eq:SCN2} is computed using \eqref{eq:number_samples} as  
\begin{align*}
	\min\Big\{N\in\mathbb{N}\mid&\sum_{i=0}^{5} \dbinom{N}{i}(1.389\times10^{-5})^{\;i}
	(0.99999)^{N-i} \\
	&\leq 0.005 \Big\} =1018779.
\end{align*}
By enforcing $\hat{M}=0.005$, the required number of samples for the approximation of the expected value in \eqref{eq:SCN2} is computed as $\hat{N}=4445$. Now, we solve the scenario problem SCP\textsubscript{w} with the acquired $N$ and $\hat{N}$ which gives us the optimal objective function $\mathcal{K}^*(\mathcal{W}_{N})$ as $-0.0761$. 

According to Theorem~\ref{theo:peyman}, since $\mathcal{K}^*(\mathcal{W}_{N})+\epsilon = -0.0462\leq 0$, one has
\begin{align*}
	\pr_w(\mathcal{S}\models_{\mathcal{T}_h} \Psi)\geq 1-\rho=0.9, 
\end{align*}
with a confidence of at least $1-\beta-\beta_{s}=0.99$. The acquired values for $\lambda$ and $c$ are $18.7479$, and $0.2891$, respectively. The barrier certificate constructed from solving SCP\textsubscript{w} is represented as:
\begin{align*}
	\hat{\mathrm{B}}(b,x) = \;0.0872x^2-2.1528x+11.9027.
\end{align*}
As discussed in Remark~\ref{rem:barrier}, the constructed barrier certificate from solving the SCP\textsubscript{w} satisfies the conditions \eqref{eq:bar1}-\eqref{eq:bar3} with a confidence of at least $1-\beta-\beta_s$. The constructed barrier certificate is illustrated in
Fig.~\ref{fig:Barrier_ver1}. As seen in Fig.~\ref{fig:Barrier_ver1}, conditions \eqref{eq:bar1} and \eqref{eq:bar2} are satisfied. Satisfaction of condition \eqref{eq:bar3} is also illustrated in Fig.~\ref{fig:Barrier_ver2}. 

\section{Conclusion}
\label{sec:ref}
In this paper, we proposed an approach to formally verify the safety of discrete-time continuous-space stochastic systems based on data randomly collected from the state space. We first formulated a barrier-based safety problem as a robust convex problem (RCP). Since solving the acquired RCP was not possible due to the unknown model that appeared in one of the constraints of the RCP, we provided a scenario convex problem (SCP) corresponding to the original RCP by employing a finite number of data collected from trajectories of the system. We then related the optimizer of the SCP to that of the RCP, and consequently, provided a safety guarantee over the unknown stochastic system with a priori guaranteed confidence. Finally, we applied our results to a room temperature system with unknown nonlinear dynamics. Formal controller synthesis for unknown discrete-time stochastic systems via data-driven construction of control barrier certificates is under investigation as a future work.

\bibliography{biblio}            
                                                   
\end{document}